\documentclass[aps,showpacs,preprintnumbers,amsmath,amssymb,nofootinbib]{revtex4}

\usepackage{graphicx}

\def\be{\begin{equation}}
\def\ee{\end{equation}}
\def\bea{\begin{eqnarray}}
\def\eea{\end{eqnarray}}

\def \T {\bar{T}}
\def \l {\lambda}
\def \k {\kappa}
\def \ra {\rightarrow}

\begin{document}


\title{Quark-Hadron Phase Transitions in Brane-World Cosmologies}

\author{Giuseppe De Risi$^{1,2}$}
\email{giuseppe.derisi@port.ac.uk}

\author{Tiberiu Harko$^3$}
\email{harko@hkucc.hku.hk}

\author{Francisco S.~N.~Lobo$^{1,4}$}
\email{francisco.lobo@port.ac.uk}

\author{Chun Shing Jason Pun$^3$}
\email{jcspun@hkucc.hku.hk}

\affiliation{$^1$Institute of Cosmology \& Gravitation,
             University of Portsmouth, Portsmouth PO1 2EG, UK}

\affiliation{$^2$Istituto Nazionale di Fisica Nucleare, 00186,
Roma, Italy}

\affiliation{$^3$Department of Physics and Center for Theoretical
and Computational Physics, The University of Hong Kong, Pok Fu Lam
Road, Hong Kong}

\affiliation{$^4$Centro de Astronomia e Astrof\'{\i}sica da
             Universidade de Lisboa, Campo Grande, Ed. C8 1749-016 Lisboa,
             Portugal}

\date{\today}

\begin{abstract}

When the universe was about 10 $\mu $seconds old, a first order
cosmological quark - hadron phase transition occurred at a
critical temperature of around 200 MeV. In this work, we study the
quark-hadron phase transition in the context of brane-world
cosmologies, in which our Universe is a three-brane embedded in a
five-dimensional bulk, and within an effective model of QCD. We
analyze the evolution of the physical quantities, relevant for the
physical description of the early universe, namely, the energy
density, temperature and scale factor, before, during, and after the
phase transition. To study the cosmological dynamics and evolution
we use both analytical and numerical methods. In particular, due
to the high energy density in the early Universe, we consider in
detail the specific brane world model case of neglecting the terms
linearly proportional to the energy density with respect to the
quadratic terms. A small brane tension and a high value of the
dark radiation term tend to decrease the effective temperature of
the quark-gluon plasma and of the hadronic fluid, respectively,
and to significantly accelerate the transition to a pure hadronic
phase. By assuming that the phase transition may be described by
an effective nucleation theory, we also consider the case where
the Universe evolved through a mixed phase with a small initial
supercooling and monotonically growing hadronic bubbles.

\end{abstract}


\maketitle


\section{Introduction}

The possibility that our $4D$ universe may be viewed as a brane
hypersurface embedded in a higher dimensional bulk space has
attracted considerable interest lately. A scenario with an
infinite fifth dimension in the presence of a brane can generate a
theory of gravity which mimics purely four-dimensional gravity,
both with respect to the classical gravitational potential and
with respect to gravitational radiation \cite{RS99a}. The
gravitational self-couplings are not significantly modified in
this model. This result has been obtained from the study of a
single 3-brane embedded in five dimensions, with the $5D$ metric
given by $ds^{2}=e^{-f(y)}\eta _{\mu \nu }dx^{\mu }dx^{\nu
}+dy^{2}$, which can produce a large hierarchy between the scale
of particle physics and gravity due to the appearance of the warp
factor \cite{RS99a}. Even if the fifth dimension is
uncompactified, standard $4D$ gravity is reproduced on the brane.
In contrast to the compactified case, this follows because the
near-brane geometry traps the massless graviton. Hence this model
allows the presence of large or even infinite non-compact extra
dimensions. Our brane is identified to a domain wall in a
5-dimensional anti-de Sitter space-time.

The effective gravitational field equations on the brane-world, in
which all the matter forces except gravity are confined on the
3-brane in a 5-dimensional space-time with $Z_2$-symmetry have
been obtained, by using an elegant geometric approach, in
\cite{SMS00,SSM00}. The correct signature for gravity is provided
by the brane with positive tension. If the bulk space-time is
exactly anti-de Sitter, then generically the matter on the brane
is required to be spatially homogeneous. The electric part of the
5-dimensional Weyl tensor $E_{IJ}$ gives the leading order
corrections to the conventional Einstein equations on the brane.
This implies a modification of the basic equations describing the
cosmological and astrophysical dynamics, which has been
extensively considered \cite{all2}. For reviews of the dynamics
and geometry of the brane Universes, as well as for the
discussions of the cosmological implications see \cite{Ma01}.

According to standard cosmology, as it expanded and cooled, the
early Universe is expected to have undergone a series of
symmetry-breaking phase transitions, at which topological defects
may have formed. Phase transitions are labelled first or second
order, according to whether the position of the vacuum state in
field space changes discontinuously or continuously, as the
critical temperature is crossed. A first order phase transition
proceeds by bubble nucleation and expansion. When at least $(4-n)$
of these bubble collide, for $n=0,1,2$, an $n$-dimensional
topological defect may form in the region between them
\cite{Kajantie:1986hq}.
Recent lattice QCD calculations for two quark flavors suggest that
QCD makes a transition at a temperature of $T_{c}\sim 150$ MeV
\cite{TaBo07}. This phase transition, which may have occurred in
the early Universe, could lead to the formation of relic
quark-gluon plasma objects, which still survive today.

A first order quark-hadron phase transition in the expanding
Universe can be described generically as follows
\cite{Kajantie:1986hq}. As the color deconfined quark-gluon plasma
cools below the critical temperature $T_{c}\approx 150$ MeV, it
becomes energetically favorable to form color confined hadrons
(primarily pions and a tiny amount of neutrons and protons, due to
the conserved net baryon number). However, the new phase does not
show up immediately. As is characteristic for a first order phase
transition, some supercooling is needed to overcome the energy
expense of forming the surface of the bubble and the new hadron
phase. When a hadron bubble is nucleated, latent heat is released,
and a spherical shock wave expands into the surrounding
supercooled quark-gluon plasma. This reheats the plasma to the
critical temperature, preventing further nucleation in a region
passed by one or more shock fronts. Generally, bubble growth is
described by deflagrations, with a shock front preceding the
actual transition front. The nucleation stops when the whole
Universe has reheated to $T_{c}$. This part of the phase
transition passes very fast, in about $0.05$ $\mu $s, during which
the cosmic expansion is totally negligible. After that, the hadron
bubbles grow at the expense of the quark phase and eventually
percolate or coalesce. The transition ends when all quark-gluon
plasma has been converted to hadrons, neglecting possible quark
nugget production. The physics of the quark-hadron phase transition, as well as the cosmological implications of this process have been extensively discussed in the framework of general relativistic cosmology in \cite{IgKaKuLa94} - \cite{IgSc01}

In the context of brane-world scenarios, the Friedmann equation
contains deviations to the $4D$ case, which results in an
increased expansion in early times. This in general has important
effects on the cosmological evolution, and in particular on
cosmological phase transitions.  First order phase transitions
have also been considered, in the framework of the brane-world
model, in \cite{DaVe01}. Due to the effects coming from the
extra-dimensions, phase transitions require a higher nucleation
rate to complete, and baryogenesis and particle abundances could
be suppressed. The evolution of topological defects is also
affected, but the increased expansion cannot solve the monopole
and domain wall problems.

In this work, we consider the quark-hadron phase transition in the
brane-world scenario. By assuming that the phase transition is of
the first order, we study in detail the evolution of the relevant
cosmological parameters (energy density, temperature, scale
factor, etc) of the quark-gluon and hadron phases, and the phase
transition itself. It is important to emphasize that in the early
universe the energy density is high, so that one may neglect the
terms linearly proportional to the energy density with respect to
the quadratic terms. This is carried out in detail in this work.
An important parameter to describe the phase transition is the
hadron fraction, whose time evolution describes the conversion
process. We also consider the effect of the dark radiation term on
the phase transition. The brane world effects (the quadratic
corrections to the matter energy-momentum tensor, described by the
numerical value of the brane tension, and the dark radiation term)
lead to an overall decrease of the temperature of the very early
universe, and accelerate the transition to the pure hadronic
phase.

This paper is organized in the following manner. In Section
\ref{SecII}, we briefly outline, for self-completeness and
self-consistency, the field equations in brane-world models and
the basics of the quark-hadron phase transition. In the latter, we
lay down the equations of state and the relevant physical
quantities that are analyzed in the remaining Sections. In Section
\ref{SecIII} we analyze in detail the quark-hadron phase
transition. In Section \ref{SecIV}, we consider bubble nucleation
in the brane-world scenario, by assuming that the phase transition
may be described by an effective nucleation theory. We discuss and
summarize our results in Section \ref{SecV}. In the present paper
we use a system of units so that the speed of light is $c=1$.


\section{Geometry, Brane-World Field Equations, Equations of State and
Consequences}\label{SecII}

\subsection{The field equations in the brane-world models}

In the brane-world models the effective
four-dimensional gravitational field equations on the brane take the
form \cite{SMS00,SSM00}:
\begin{equation}
G_{\mu\nu} = - \Lambda g_{\mu\nu} + k_4^2 T_{\mu\nu}+k_5^4 S_{\mu
\nu} - E_{\mu\nu},
\end{equation}
where $\Lambda =k_5^2(\Lambda_5+k_5^2 \lambda^2/6)/2$ and
$k_4^2=k_5^4\lambda/6$. $T_{\mu\nu}$ is the matter energy-momentum
tensor on the brane and $T=T^\mu{}_\mu$ is the trace of the
energy-momentum tensor. The Einstein equation in the bulk also implies the conservation of the
energy momentum tensor of the matter on the brane \cite{SSM00}. The first correction term relative to
Einstein's general relativity is the inclusion of a quadratic term
$S_{\mu\nu}$ in the energy-momentum tensor, arising from the
extrinsic curvature terms in the projected Einstein tensor, and is
given by
\begin{equation}
S_{\mu\nu} = \frac1{12} T T_{\mu\nu} - \frac14 T_\mu{}^\alpha
T_{\nu\alpha} + \frac1{24} g_{\mu\nu} \left( 3 T^{\alpha\beta}
T_{\alpha\beta} - T^2 \right)\,.
\end{equation}
The second correction term, $E_{\mu\nu}$, is the projection of the
5-dimensional Weyl tensor, $C_{ABCD}$, onto the brane, and is
defined as
$E_{\mu\nu}=\delta_\mu^A\,\delta_\nu^C\;C_{ABCD}\,n^Bn^D$, and
encompasses nonlocal bulk effects. The only general known property
of this nonlocal term is that it is traceless, i.e.,
$E^{\mu}{}_{\mu}=0$. The symmetry properties of $E_{\mu \nu }$ imply that in general we
can decompose it irreducibly with respect to a chosen 4-velocity
field $u^{\mu }$ as
\begin{equation}
E_{\mu \nu }=-\frac{6}{\lambda k_{4}^{2}}\left[ {\cal U}\left( u_{\mu
}u_{\nu }+\frac{1}{3}h_{\mu \nu }\right) +{\cal P}_{\mu \nu }+2{\cal Q}%
_{(\mu }u_{\nu )}\right] ,  \label{WT}
\end{equation}
where $h_{\mu \nu }=g_{\mu \nu }+u_{\mu }u_{\nu }$ projects orthogonal
to $u^{\mu }$, ${\cal U}$ is a scalar, ${\cal Q}_{\mu }$ a spatial vector
and ${\cal P}_{\mu \nu }$ a spatial, symmetric and trace-free
tensor, respectively. For homogeneous and isotropic Friedmann-Robertson-Walker (FRW) type cosmological models ${\cal Q}_{\mu }={\cal P}_{\mu \nu }=0$
\cite{Ma01}, and hence the only non-zero contribution from the
5-dimensional Weyl tensor from the bulk is given by the scalar
term ${\cal U}$.

There are several constraints which have been obtained for the
brane tension $\lambda $. Thus from big bang nucleosynthesis
constraints it follows $ \lambda \geq 1$ MeV$^{4}$ \cite{GeMa01}.
A much stronger bound for $\lambda $ arises from null results of
submillimeter tests of Newton's law, giving $\lambda \geq 10^{8}$
GeV$^{4}$ \cite{MaWaBaHe00}. An astrophysical lower limit on
$\lambda $, which is independent of the Newton law and the
cosmological limits has been derived in \cite{GeMa01}, leading to
$\lambda
>5\times 10^{8}$ MeV$^{4}$.\

In this work, we assume that the space-time geometry is the flat
FRW metric, given by
\begin{equation}
ds^{2}=-dt^{2}+a^{2}(t)\left( dx^{2}+dy^{2}+dz^{2}\right).
\label{R6}
\end{equation}
For the matter energy-momentum tensor on the brane we restrict our
analysis to the case of the perfect fluid energy-momentum tensor,
\begin{equation}
T^{\mu \nu }=(\rho +p)u^{\mu }u^{\nu }+pg^{\mu \nu },
\end{equation}
where $\rho $ and $p$ are the energy density and isotropic pressure of the cosmological fluid on the brane, respectively. 

Thus, the gravitational field equations, corresponding to the line
element (\ref{R6}) become
\begin{eqnarray}
3\frac{\dot{a}^{2}}{a^{2}} &=&\Lambda +k_{4}^{2}\rho +\frac{k_{4}^{2}}{%
2\lambda }\rho ^{2}+\frac{6}{\lambda k_{4}^{2}}{\cal U},  \label{dH} \\
2\frac{\ddot{a}}{a}+\frac{\dot{a}^{2}}{a^{2}} &=&\Lambda -k_{4}^{2}p-\frac{%
k_{4}^{2}}{\lambda }\rho p-\frac{k_{4}^{2}}{2\lambda }\rho ^{2}-\frac{2}{%
\lambda k_{4}^{2}}{\cal U},  \label{dVHi} \\
\dot{\rho}+3(\rho +p)\frac{\dot{a}}{a} &=&0,  \label{drho} \\
\dot{{\cal U}}+4\frac{\dot{a}}{a}{\cal U} &=&0,  \label{u}
\end{eqnarray}

Eq. (\ref{u}) can be immediately integrated to give the following general
expression for the ``dark energy'' ${\cal U}$:
\begin{equation}
{\cal U}=\frac{{\cal U}_{0}}{a^{4}},
\end{equation}
with ${\cal U}_{0}>0$ a constant of integration.

\subsection{Quark-hadron phase transition}

In this Section, we outline the relevant physical quantities of
the quark-hadron phase transition, which will be applied in the
following sections in the context of the brane-world scenario.
Note that the scale of the cosmological QCD transition is given by
the Hubble radius $R_{H}$ at the transition, which is $R_{H}\sim
m_{Pl}/T_{c}^{2}\sim 10\,$km, where $T_c$ is the critical
temperature. The mass inside the Hubble volume is $\sim
1\,M_{\odot}$. The expansion time scale is $10^{-5}$ s, which
should be compared with the timescale of QCD, $1$ fm/c$\approx
10^{-23}$ s. Even the rate of the weak interactions exceeds the
Hubble rate by a factor of $10^{7}$. Therefore, in this phase
photons, leptons, quarks and gluons (or pions) are lightly coupled
and may be described as a single, adiabatically expanding fluid
\cite{ChMa96}.


In order to study the quark-hadron phase transition it is
necessary to specify the equation of state of the matter, in both
quark and hadron state. Giving an equation of state is equivalent
to give the pressure as a function of the temperature $T$ and
chemical potential $\mu $. At high temperatures the quark chemical potentials are equal,
because weak interactions keep them in chemical equilibrium, and
the chemical potentials for leptons are assumed to vanish. Thus
the chemical potential for a baryon is defined by $\mu _{B}=3\mu
_{q}$. The baryon number density of an ideal Fermi gas of three
quark flavors is given by $n_{B}\approx T^{2}\mu _{B}/3$, leading
to $\mu _{B}/T\sim 10^{-9}$ at $T>T_{c}$. At low temperatures $\mu
_{B}/T\sim 10^{-2}$. Therefore the assumption of a vanishing
chemical potential at the phase transition temperature in both
quark and hadron phase represents an excellent approximation for
the study of the equation of state of the cosmological matter in
the early Universe. In addition to the strongly interacting matter,
we assume that in each phase there are present leptons and
relativistic photons, satisfying equations of state similar to
that of hadronic matter \cite{Kajantie:1986hq}.

The equation of state of the matter in the quark phase can
generally be given in the form
\begin{equation}
\rho _{q}=3a_{q}T^{4}+V(T), \qquad p_{q}=a_{q}T^{4}-V(T),
\label{eqq}
\end{equation}
where $a_{q}=\left( \pi ^{2}/90\right) g_{q}$, with
$g_{q}=16+(21/2)N_{F}+14.25=51.25$ and $N_{F}=2$. $V\left(T\right)
$ is the self-interaction potential. For $V$ we adopt the
expression \cite {BoCoMa00}
\begin{equation}
V\left( T\right) =B+\gamma _{T}T^{2}-\alpha _{T}T^{4},  \label{V}
\end{equation}
where $B$ is the bag constant, $\alpha _{T}=7\pi ^{2}/20$, and
$\gamma_{T}=m_{s}^{2}/4$, with $m_{s}$ the mass of the strange
quark in the range $m_{s}\in \left( 60-200\right) $ MeV. The form
of the potential $V$ corresponds to a physical model in which the
quark fields are interacting with a chiral field formed with the
$\pi $ meson field and a scalar field. If the temperature effects
can be ignored, the equation of state in the quark phase takes the
form of the MIT bag model equation of state, $p_{q}=(\rho
_{q}-4B)/3$. Results obtained in low energy hadron spectroscopy,
heavy ion collisions and phenomenological fits of light hadron
properties give $B^{1/4}$ between $100$ and $200$ MeV
\cite{LePa92}.

In the hadron phase we take the cosmological fluid to consist of
an ideal gas of massless pions and of nucleons described by the
Maxwell-Boltzmann statistics, with energy density $\rho _{h}$ and
pressure $p_{h}$, respectively. The equation of state can be
approximated by
\begin{equation}
p_{h}\left( T\right) =\frac{1}{3}\rho _{h}\left( T\right) =a_{\pi }T^{4},
\label{eqh}
\end{equation}
where $a_{\pi }=\left( \pi ^{2}/90\right) g_{h}$ and
$g_{h}=17.25$.

The critical temperature $T_{c}$ is defined by the condition
$p_{q}\left( T_{c}\right) =p_{h}\left( T_{c}\right) $
\cite{Kajantie:1986hq}, and is given, in the present model, by
\begin{equation}
T_{c}=\sqrt{\frac{\gamma _{T}+\sqrt{\gamma _{T}^{2}+4\left(
a_{q}+\alpha _{T}-a_{\pi }\right) B}}{2\left( a_{q}+\alpha
_{T}-a_{\pi }\right) }}. \label{Tc}
\end{equation}

For $m_{s}=200$ MeV and $B^{1/4}=200$ MeV the transition
temperature is of the order $T_{c}\approx 125$ MeV. Since the
phase transition is of first order, all the physical quantities,
like the energy density, pressure and entropy exhibit
discontinuities across the critical curve. The ratio of the quark and hadron energy densities at the critical temperature, $\rho _{q}\left( T_{c}\right) /\rho _{h}\left(
T_{c}\right) $, is of the order of $3.62$ for $m_{s}=200$ MeV and
$B^{1/4}=200$ MeV. If the temperature effects in the
self-interaction potential $V$ are neglected, $\alpha _{T}=\gamma
_{T}\approx 0$, then we obtain the well-known
relation between the critical temperature and the bag constant,
$B=\left( g_{q}-g_{h}\right) \pi ^{2}T_{c}^{4}/90$
\cite{Kajantie:1986hq}.



\section{Dynamics of the brane Universe during the quark-hadron phase
transition}\label{SecIII}

The quantities to be traced through the quark-hadron phase
transition in the brane-world cosmological scenario are the energy
density $\rho $, the temperature $T$ and the scale factor $a$.
These quantities are determined by the gravitational field
equations (\ref{dH}) and (\ref{drho}) and by the equations of
state (\ref{eqq}), (\ref{V}) and (\ref{eqh}). We shall consider
now the evolution of the brane-world before, during and after the
phase transition.

Before the phase transition, $T>T_{c}$ the brane-world is in the
quark phase. With the use of the equations of state of the quark
matter, and the Bianchi identity on the brane, Eq. (\ref{drho})
can be written in the form
\begin{equation}
\frac{\dot{a}}{a}=-\frac{3a_{q}-\alpha
_{T}}{3a_{q}}\frac{\dot{T}}{T}-\frac{1}{6}\frac{\gamma
_{T}}{a_{q}}\frac{\dot{T}}{T^{3}},  \label{rdot}
\end{equation}
and can be integrated to give the following scale
factor-temperature relation:
\begin{equation}
a(T)=a_{0}T^{\frac{\alpha _{T-3a_{q}}}{3a_{q}}}\exp \left(
\frac{1}{12}\frac{\gamma _{T}}{a_{q}}\frac{1}{T^{2}}\right) ,
\end{equation}
where $a_{0}$ is a constant of integration.

With the use of Eq. (\ref{rdot}), from the gravitational field
equations we obtain an expression, describing the evolution of the
temperature of the brane Universe in the quark phase, given by
\begin{equation}
\frac{dT}{dt}=-\frac{T^{3}}{\sqrt{3}}\frac{\sqrt{
A_{0}T^{8}+A_{1}T^{6}+A_{2}T^{4}+A_{3}T^{2}+A_{4}T^{4A_{5}}\exp
\left( -2A_{6}\frac{1}{T^{2}}\right) +A_{7}}}{A_{5}T^{2}+A_{6}},
\end{equation}
where we have denoted
\begin{eqnarray}
A_{0}=k_{4}^{2}\left( 3a_{q}-\alpha _{T}\right) ^{2}/2\lambda ,
\quad A_{1}=k_{4}^{2}\left( 3a_{q}-\alpha _{T}\right) \gamma
_{T}/\lambda ,
\quad A_{2}=k_{4}^{2}\left[ \gamma _{T}^{2}/2\lambda +\left(
3a_{q}-\alpha
_{T}\right) \left( 1+B/\lambda \right) \right] , \\
A_{3}=k_{4}^{2}\left( 1+B/\lambda \right) \gamma _{T},
\quad A_{4}=6{\cal U} _{0}/\lambda k_{4}^{2}a_{0}^{4},A_{5}=\left(
3a_{q}-\alpha _{T}\right) /3a_{q},
\quad A_{6}=\gamma _{T}/6a_{q},
\quad A_{7}=\Lambda +k_{4}^{2}B\left( 1+B/2\lambda \right) .
\end{eqnarray}

The variation of the temperature as a function of the parameter
$\tau =k_4t$ in the quark matter filled brane-world is
represented, in the case of a vanishing cosmological constant
$\Lambda =0$, for different values of the brane tension $\lambda $
and for a fixed numerical value of the constant $A_{4}=6{\cal U}
_{0}/\lambda k_{4}^{2}a_{0}^{4}$, in Fig. 1.

\vspace{0.1in}
\begin{figure}[h]
\includegraphics{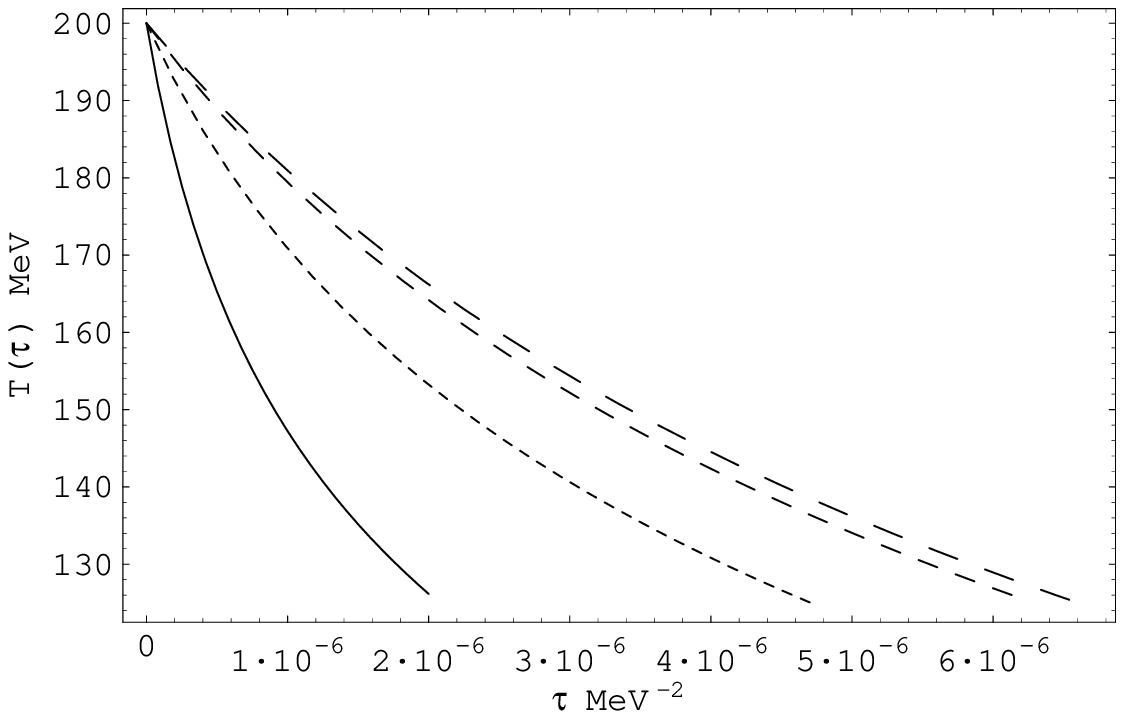}
\caption{ Variation of the temperature of the quark fluid on the
brane as a function of $\tau =k_4t$ for different values of the
brane tension $\lambda $: $\lambda =5\times 10^8$ MeV$^4$ (solid
curve), $\lambda =5\times 10^9$ MeV$^4$ (dotted curve), $\lambda
=5\times 10^{10}$ MeV$^4$ (short dashed curve) and $\lambda
=5\times 10^{16}$ MeV$^4$ (long dashed curve). We have assumed
that in this phase the cosmological constant is vanishingly small.
For the bag constant we have chosen the value $B^{1/4}=200$ MeV,
while $A_4=0.01$. } \label{FIG1}
\end{figure}

The variation of the temperature of the brane in the quark phases
for different values of the constant $A_4$, corresponding to
different numerical values of the integration constant ${\cal
U}_0$, describing the effect of the dark radiation term on the
cosmological evolution, and considering a fixed value for the
brane tension $\lambda$, is represented in Fig. 2.

\vspace{0.1in}
\begin{figure}[h]
\includegraphics{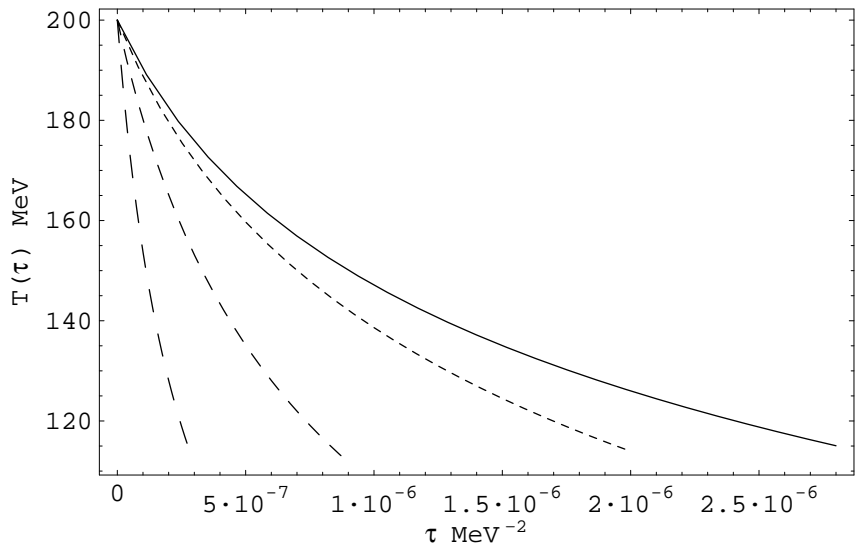}
\caption{ Variation of the temperature of the quark fluid on the
brane as a function of $\tau =k_4t$ for different values of the
dark radiation term $\cal U$: $A_4 =10^2$ (solid curve), $A_4
=10^4$ (dotted curve), $A_4 =10^5$ (short dashed curve) and $A_4
=10^6$ (long dashed curve). We have assumed that in this phase the
cosmological constant is vanishingly small. For the bag constant
and for the brane tension we have chosen the values $B^{1/4}=200$
MeV and $\lambda=5\times 10^8$ MeV$^4$, respectively.}
\label{FIG2}
\end{figure}

In order to have an analytical insight into the evolution of the
cosmological quark matter in the brane-world and on the effects of
the extra-dimensions on the phase transition, we consider the
simple case in which the temperature corrections can be neglected
in the self-interaction potential $V$. In this case $V=B={\rm
constant}$, and the equation of state of the quark matter is given
by the bag model equation of state, $p_{q}=\left( \rho
_{q}-4B\right) /3$. Thus, Eq. (\ref{drho}) can immediately be
integrated to give the following scale factor-temperature
relation:
\begin{equation}
a(T)=\frac{a_{0}}{T},
\end{equation}
where $a_{0}$ is a constant of integration. Since in the initial
stages of the evolution of the Universe the density of the matter
is very high, we suppose that in Eq. (\ref{dH}) we can neglect the
term proportional to the energy density with respect to the term
containing $\rho ^{2}$. The contributions of the cosmological
constant and of the dark energy are also negligible small.
Therefore the evolution of the quark phase of the brane-world is
described by the equation
\begin{equation}
\frac{\dot{a}}{a}\approx \frac{k_{4}}{\sqrt{6\lambda }}\rho .
\end{equation}

Hence the time dependence of the temperature can be obtained from
the equation
\begin{equation}
\frac{dT}{dt}\approx -\frac{k_{4}}{\sqrt{6\lambda }}\left(
3a_{q}T^{5}+BT\right),
\label{dT_quark}
\end{equation}
with the general solution given by
\begin{equation}
T(t)\approx \frac{B^{1/4}C^{1/4}\exp \left(
-\frac{k_{4}B}{\sqrt{6\lambda }} t\right) }{\left[ 1-3a_{q}C\exp
\left( -\frac{4k_{4}B}{\sqrt{6\lambda }} t\right) \right] ^{1/4}}.
\label{T(t)_quark}
\end{equation}
The integration constant $C$ is related to the temperature $T_{0}$
of the quark matter at the time $t_{0}$ by means of the relation
\begin{equation}
C=\frac{T_{0}^{4}\exp \left( \frac{4k_{4}B}{\sqrt{6\lambda
}}t_{0}\right)}{\left( 3a_{q}T_{0}^{4}+B\right) ^{1/4}},
\end{equation}
leading to
\begin{equation}
T(t)\approx \frac{B^{1/4}T_{0}\exp \left[
-\frac{k_{4}B}{\sqrt{6\lambda}} \left( t-t_{0}\right) \right]
}{\left\{ B+3a_{q}T_{0}^{4}-3a_{q}T_{0}^{4}\exp \left[
-\frac{4k_{4}B}{\sqrt{6\lambda}}\left( t-t_{0}\right) \right]
\right\} ^{1/4}}.
\end{equation}

During the phase transition, the temperature and the pressure are
constants, $T=T_{c}$ and $p=p_{c}$, respectively. The entropy
$S=sa^{3}$ and the enthalpy $W=\left( \rho +p\right) a^{3}$ are
conserved quantities. During the phase transition $\rho \left(
t\right) $ decreases from $\rho _{q}\left( T_{c}\right) \equiv
\rho _{Q}$ to $\rho _{h}\left( T_{c}\right) \equiv \rho _{H}$. For
phase transition temperature of $T_{c}=125$ MeV we have $\rho
_{Q}\approx 5\times 10^{9}$ MeV$^{4}$ and $\rho _{H}\approx
1.38\times 10^{9} $ MeV$^{4}$, respectively. For the same value of
the temperature the value of the pressure of the cosmological
fluid during the phase transition is $p_{c}\approx 4.6\times
10^{8}$ MeV$^{4}$. It is convenient, following
\cite{Kajantie:1986hq}, to replace $\rho \left( t\right) $ by
$h(t)$, the volume fraction of matter in the hadron phase, by
defining
\begin{equation}
\rho \left( t\right) =\rho _{H}h(t)+\rho _{Q}\left[ 1-h(t)\right]
=\rho _{Q} \left[ 1+nh(t)\right] ,
\end{equation}
where we denoted $n=\left( \rho _{H}-\rho _{Q}\right) /\rho _{Q}$.
At the beginning of the phase transition $h(t_{c})=0$, where
$t_{c}$ is the time corresponding to the beginning of the phase
transition, and $\rho \left( t_{c}\right) \equiv \rho _{Q}$, while
at the end of the transition $h\left( t_{h}\right) =1$, where
$t_{h}$ is the time at which the phase transition ends,
corresponding to $\rho \left( t_{h}\right) \equiv \rho _{H}$. For
$t>t_{h}$ the Universe enters in the hadronic phase.

From Eq. (\ref{drho}) we obtain
\begin{equation}
\frac{\dot{a}}{a}=-\frac{1}{3}\frac{\left( \rho _{H}-\rho
_{Q}\right) \dot{h} }{\rho _{Q}+p_{c}+\left( \rho _{H}-\rho
_{Q}\right) h}=-\frac{1}{3}\frac{r \dot{h}}{1+rh},
\end{equation}
where we denoted $r=\left( \rho _{H}-\rho _{Q}\right) /\left( \rho
_{Q}+p_{c}\right)$. The above equation immediately leads to
\begin{equation}
a(t)=a\left( t_{c}\right) \left[ 1+rh(t)\right] ^{-1/3},
\end{equation}
where we have used the initial condition $h\left(t_{c}\right)=0$.
The evolution of the fraction of the matter in the hadron phase is
described by the equation
\begin{equation}
\frac{dh}{dt}=-\sqrt{3}\left( h+\frac{1}{r}\right)\sqrt{\Lambda
+k_{4}^{2}\rho _{Q}+\frac{k_{4}^{2}\rho _{Q}^{2}}{2\lambda
}+k_{4}^{2}n\rho _{Q}\left( 1+\frac{\rho _{Q}}{\lambda }\right)
h+\frac{k_{4}^{2}\rho _{Q}^{2}n^{2}}{2\lambda }h^{2}+\frac{6{\cal
U}_{0}}{k_{4}^{2}\lambda a^{4}\left( t_{c}\right) }\left(
rh+1\right) ^{4/3}}.
\end{equation}

It is also convenient to plot the variation of the hadron fraction
as a function of the different parameters. The variation of the
hadron fraction $h$ as a function of the parameter $\tau =k_4t$ is
represented, for different values of the brane tension and for a
fixed value of ${\cal U} _0$ in Fig. 3.
%
\begin{figure}[h]
\includegraphics{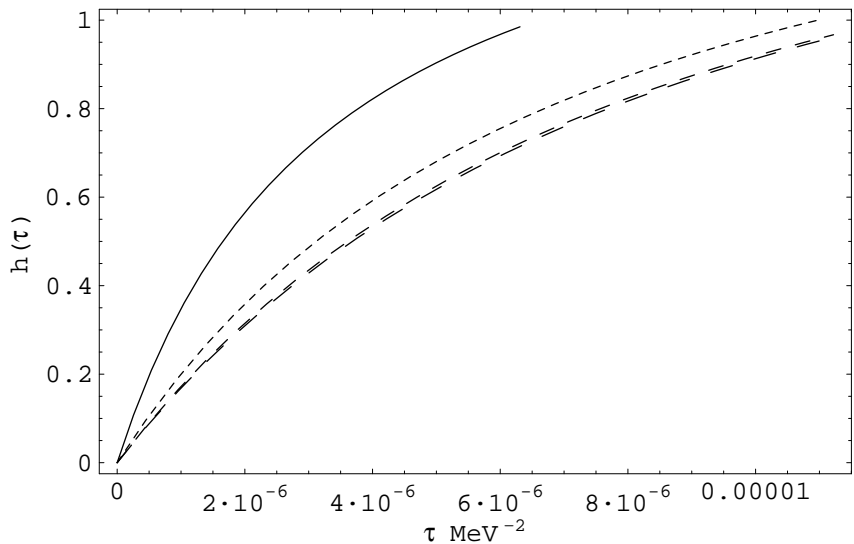}
\caption{ Variation of the hadron fraction $h$ as a function of
the parameter $\tau =k_4t$ during the quark-hadron phase
transition on the brane, for different values of the brane
tension: $\lambda =5\times 10^8$ MeV$^4$ (solid curve), $\lambda
=5\times 10^9$ MeV$^4$ (dotted curve), $\lambda =5\times 10^{10}$
MeV$^4$ (short dashed curve) and $\lambda =5\times 10^{12}$
MeV$^4$ (long dashed curve). We have assumed that the effect of
the cosmological constant can be neglected. The temperature during
the phase transition is $T_c=125$ MeV. For the bag constant we
have chosen the value $B^{1/4}=200$ MeV, while $A_4=0.01$.}
\label{FIG3}
\end{figure}

The time variation of the hadron fraction during the phase
transition for a fixed brane tension $\lambda $ and for different
values of the integration constant ${\cal U}_0$ is represented in
Fig. 4.
\vspace{0.1in}
\begin{figure}[h]
\includegraphics{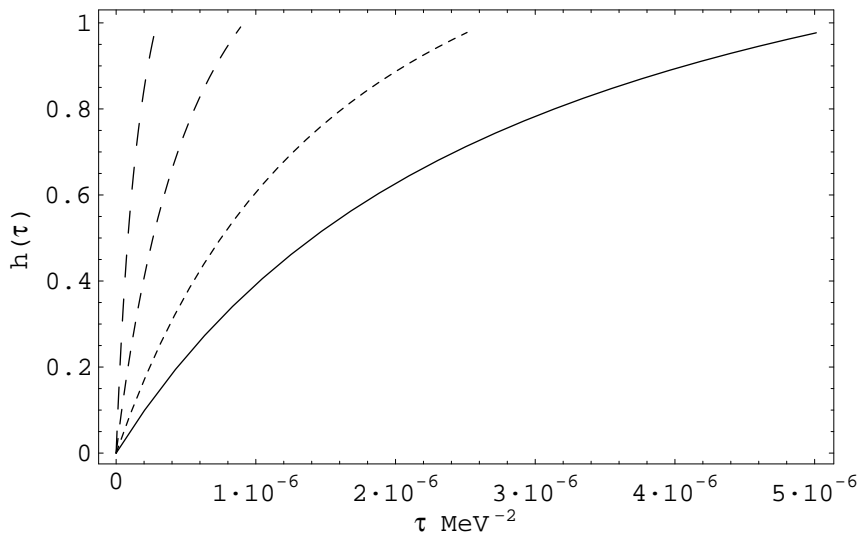}
\caption{ Variation of the hadron fraction $h$ as a function of
the parameter $\tau =k_4t$ during the quark-hadron phase
transition on the brane, for different values of the dark
radiation term $u_0=6{\cal U}_0/K_4^2\lambda a^4\left(t_c\right)$:
$u_0 =10^{10}$ MeV$^2$ (solid curve), $u_0=10^{11}$ MeV$^2$
(dotted curve), $u_0=10^{12}$ MeV$^2$ (short dashed curve) and
$u_0=10^{13}$ MeV$^2$ (long dashed curve). We have assumed that
the effect of the cosmological constant can be neglected. The
temperature during the phase transition is $T_c=125$ MeV. For the
bag constant we have chosen the value $B^{1/4}=200$ MeV, while
$\lambda =5\times 10^8$.} \label{FIG4}
\end{figure}

If the quadratic term in the energy density, describing the
effects of the extra-dimensions, dominates the evolution of the
Universe, and by neglecting the dark energy contribution, the
evolution of the hadron fraction during the phase transition is
described by the equation
\begin{equation}
\frac{dh}{dt}=-\frac{3k_{4}\rho _{Q}}{r\sqrt{6\lambda }}\left( 1+nh\right)
\left( 1+rh\right) ,
\end{equation}
with the general solution given by
\begin{equation}
h(t)=\frac{\exp \left[ -\frac{3\left( n-r\right) k_{4}\rho
_{Q}}{r\sqrt{6\lambda }}\left( t-t_{c}\right) \right] -1}{n-r\exp
\left[ -\frac{3\left( n-r\right) k_{4}\rho _{Q}}{r\sqrt{6\lambda
}}\left( t-t_{c}\right) \right] },
\end{equation}
where we have also used the initial condition $h\left(
t_{c}\right)=0$.

The respective evolution of the scale factor during the phase
transition for the extra-dimensions dominated evolution of the
cosmological fluid is described by the equation
\begin{equation}
a(t)=\frac{a\left( t_{c}\right) }{\left( n-r\right) ^{1/3}}\left\{
n-r\exp \left[ -\frac{3\left( n-r\right) k_{4}\rho
_{Q}}{r\sqrt{6\lambda }}\left( t-t_{c}\right) \right] \right\}
^{1/3}.
\end{equation}

The phase transition ends when $h(t)$ has increased to $1$. If the
evolution is dominated by the terms coming from the
extra-dimensions in the bulk, the time $t_{h}$ at which the phase
transition ends is
\begin{equation}
t_{h}=t_{c}+\frac{\sqrt{6\lambda }r\ln \frac{1+n}{1+r}}{3k_{4}\left(
r-n\right) \rho _{Q}}.
\end{equation}
At the end of the phase transition the scale factor of the
Universe has the value $a\left( t_{h}\right) =a\left( t_{c}\right)
\left( r+1\right) ^{-1/3}$.

Finally, after the phase transition, the energy density of the
pure hadronic matter is $\rho _{h}=3p_{h}=3a_{\pi }T^{4}$. The
Bianchi identity Eq. (\ref {drho}) gives $a(T)=a\left(
t_{h}\right) T_{c}/T$. The temperature dependence of the brane
Universe in the hadronic phase is governed by the equation
\begin{equation}
\frac{dT}{dt}=-\frac{T}{\sqrt{3}}\sqrt{\Lambda +\left( 3a_{\pi
}k_{4}^{2}+ \frac{6{\cal U}_{0}}{\lambda k_{4}^{2}a^{4}\left(
t_{h}\right) T_{c}^{4}} \right) T^{4}+\frac{9a_{\pi
}^{2}k_{4}^{2}}{2\lambda }T^{8}}.
\end{equation}

As before, it is convenient to plot the variation of the
temperature in terms of the various parameters. The variation of
the temperature of the hadron fluid filled brane Universe as a
function of the brane tension $\lambda $ is represented in Fig. 5.

\vspace{0.1in}
\begin{figure}[h]
\includegraphics{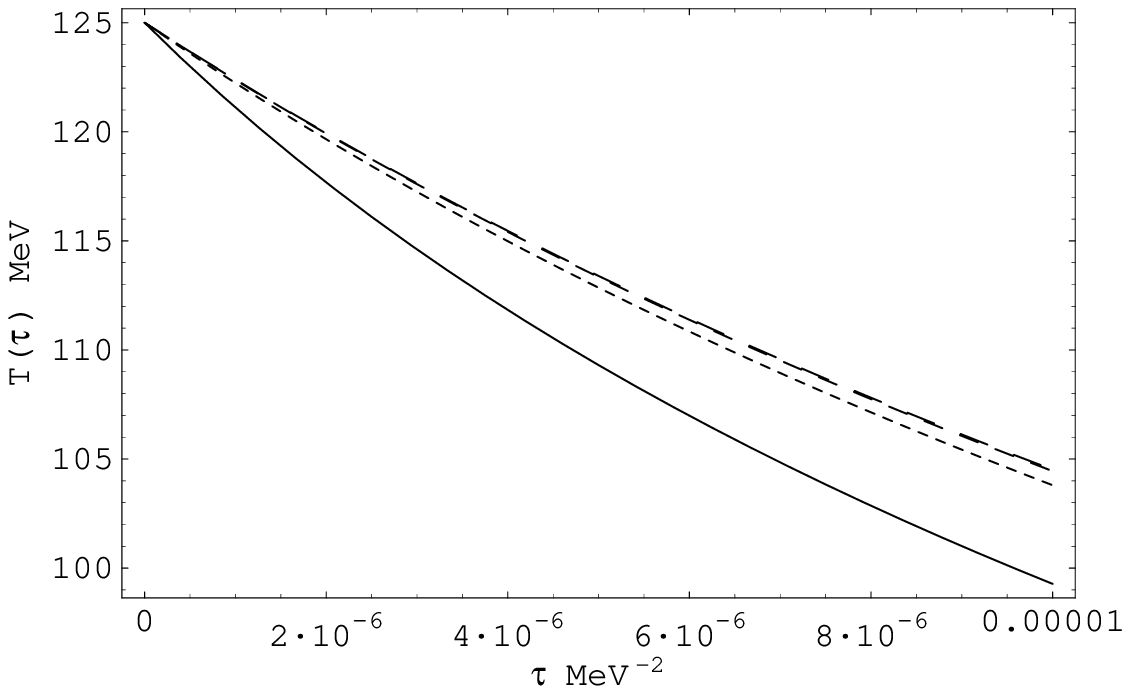}
\caption{ Variation of the temperature of the hadron fluid on the
brane, as a function of $\tau =k_4t$, for different values of the
brane tension $\lambda $: $\lambda =5\times 10^8$ MeV$^4$ (solid
curve), $\lambda =5\times 10^9$ MeV$^4$ (dotted curve), $\lambda
=5\times 10^{10}$ MeV$^4$ (short dashed curve) and $\lambda
=5\times 10^{11}$ MeV$^4$ (long dashed curve). We have assumed
that in this phase the cosmological constant is vanishingly small.
For the bag constant we have chosen the value $B^{1/4}=200$ MeV,
while the dark radiation term has been fixed so that $6{\cal
U}_{0}/\lambda k_{4}^{4}a^{4}\left( t_{h}\right) T_{c}^{4}=0.01$.
} \label{FIG5}
\end{figure}

The time variation of the temperature for the brane in the hadron
phase for different values of the dark energy coefficient ${\cal
U}_0$ is represented in Fig. 6.

\vspace{0.1in}
\begin{figure}[h]
\includegraphics{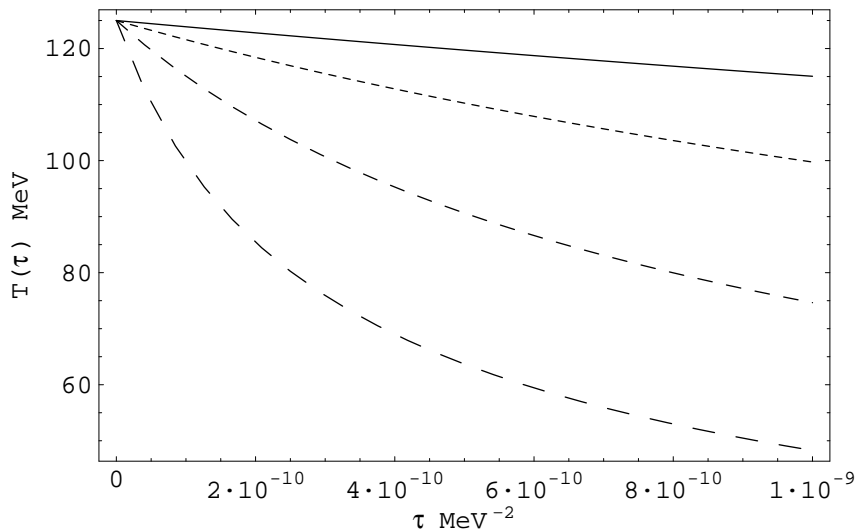}
\caption{ Variation of the temperature of the hadronic fluid on
the brane, as a function of $\tau =k_4t$, for different values of
the dark radiation term $u_0=6{\cal U}_{0}/\lambda
k_{4}^{4}a^{4}\left( t_{h}\right) T_{c}^{4}$: $u_0 =10^8$ (solid
curve), $u_0 =10^9$ (dotted curve), $u_0 =10^{10}$ (short dashed
curve) and $u_0 =10^{11}$ (long dashed curve). We have assumed
that in this phase the cosmological constant is vanishingly small.
For the bag constant and for the brane tension we have chosen the
values $B^{1/4}=200$ MeV and $\lambda=5\times 10^8$ MeV$^4$,
respectively.} \label{FIG6}
\end{figure}

If the extra-dimensional terms dominates the four-dimensional
gravitational effects, then the time variation of the temperature
of the cosmological hadron fluid is given by
\begin{equation}
T(t)\approx \left[ \frac{12k_{4}a_{\pi }}{\sqrt{6\lambda }}\left(
t-t_{h}\right) +T_{c}^{-4}\right] ^{-1/4},
\end{equation}
where we have used the initial condition $T\left(
t_{h}\right)=T_{c}$.

In the extra-dimensions dominated hadron phase the scale factor of
the Universe evolves according to the law
\begin{equation}
a(t)\approx a\left( t_{h}\right) T_{c}\left[ \frac{12k_{4}a_{\pi
}}{\sqrt{ 6\lambda }}\left( t-t_{h}\right) +T_{c}^{-4}\right]
^{1/4}.
\end{equation}


\section{Bubble nucleation in the brane-world cosmological scenarios}
\label{SecIV}

In order to describe the process of formation and evolution of
microscopic quark nuclei in the cosmological fluid on the brane
one must use nucleation theory. The goal of nucleation theory is
to compute the probability that a bubble or droplet of the $A$
phase appears in a system initially in the $B$ phase near the
critical temperature \cite{LaLi80}. Homogeneous nucleation theory
applies when the system is pure. Nucleation theory is applicable
for first-order phase transitions when the matter is not
dramatically supercooled or superheated. If substantial
supercooling or superheating is present, or if the phase
transition is second order, then the relevant dynamics is spinodal
decomposition \cite{ShMo01}.

In this context, note that in the analysis of the phase transition
considered in the previous section we do not take into account the
possibility that the quark plasma could undergo a supercooling
phase, in which it stays in the quark phase below the critical
temperature. If the supercooling is not too strong, the phase
transition can be described by an effective nucleation theory, as
mentioned above. As the temperature decreases, there is a
probability that a droplet of hadrons is nucleated from the quark
plasma. The nucleation probability density is given by
\cite{Kajantie:1986hq}
\begin{equation}
p(t) = p_0 T_c^4 \exp \left[ -\frac{w_0}{\left( 1 - \T^4
\right)^2} \right] \,, \label{nucl-prob}
\end{equation}
where $\T = T/T_c$. Once a droplet of hadrons is formed, it starts
to expand explosively \cite{Gyulassy:1983rq}, with a velocity
$v_{fr}$ smaller than the sound speed. Contemporarily, a much
quicker shock wave is generated. More and more bubbles are created
while the temperature decreases, until the shock waves collide and
re-heat the plasma to the critical temperature.

To calculate the fraction of the volume which at a time $t$ is
turned to the hadronic phase in the small supercooling scenario,
we should sum over the volumes at the time $t$ of the bubbles
which are nucleated at a previous time $t_p$, times the
probability to create a bubble at that time $t_p$, which is given
by
\begin{equation}
f(t) = \int_{t_i}^t dt_p p(t_p) \frac{4\pi}{3}\left[v_{fr}
\left( t-t_p\right)\right]^3,
\label{scfraction_t}
\end{equation}
where $t_i$ is the time at which the critical temperature is
reached, and the nucleation process started. Using Eqs.
(\ref{dT_quark})-(\ref{T(t)_quark}), we can express this integral
in terms of normalized temperature. After some algebra we get
\be f(\T) = \int_{\T}^1 d\T_p \frac{p_0 v_{fr}^3}{B^3}
\left(\frac{6 \l}{\k_4^2}\right)^2 \frac{\exp \left[
-\frac{w_0}{\left( 1 - \T_p^4 \right)^2} \right]} {\T_p \left( 3
a_q \T_p^4 + b\right)} \left\{ \log\left[\frac{ 3 a_q \T_p^4 + b
\left(\frac{\T_p}{\T} \right)^4}{3 a_q \T_p^4+b}\right]\right\}^3
\label{scfraction_T} \,, \ee
where $b \equiv B/T_c^4$.

The integrand is extremely peaked around its maximum, so the
integral is dominated by the value of the function at the maximum,
which can be found by solving the equation:
\be
\log \left[
\frac{3 a_q \T_p^4 + b \left(\frac{\T_p}{\T}\right)^4}{3 a_q
\T_p^4 + b} \right] \left[ \frac{8 w_0 \T_p^2}{\left(
1-\T_p^4\right)^3} + \frac{1}{\T_p^2} + \frac{12 a_q \T_p^2}{3 a_q
\T_p^4 + b}\right] = 3 \left( \frac{4}{\T_p^2} - \frac{12 a_q
\T_p^2}{3 a_q \T_p^4 + b} \right)\,. \label{der}
\ee
To evaluate this maximum, we note that the integrated function
falls very rapidly to zero as $\T_p \ra 1$ because of the
exponential, so the maximum must be very close to the lower end of
the interval. Thus we can set $\T_{p,max} \simeq \T + x$ with
$x/\T \ll 1$ and approximate the first derivative (\ref{der}) to
first order in $x$. Moreover, since we will assume only a small
supercooling, the final temperature $\T$ is very close to $1$, so
that the first term in the square bracket of the left hand side in
Eq.~(\ref{der}) is much larger than the other two, which are of
order $1$ and can be discarded. Eventually we obtain
\be
x(\T) = \frac{\left(1-\T^4\right)^3 \left( 6 a_q \T^4 +
b\right)}{8 w_0 b \T^3}, \label{x} \ee so that the integral
(\ref{scfraction_T}) can be approximated as \be f(\T) \simeq x(\T)
\frac{p_0 v_{fr}^3}{B^3} \left(\frac{6 \l}{\k_4^2}\right)^2
\frac{\exp \left\{ -\frac{w_0}{\left[ 1 - (\T+x(\T))^4 \right]^2}
\right\}}{(\T+x(\T)) \left[ 3 a_q (\T+x(\T))^4 + b\right]} \left\{
\log\left[\frac{3 a_q (\T+x(\T))^4 + b \left(\frac{\T+x(\T)}{\T}
\right)^4}{3 a_q (\T+x(\T))^4 + b}\right]\right\}^3.
\label{scfrac_app}
\ee

\begin{figure}[h]
\includegraphics{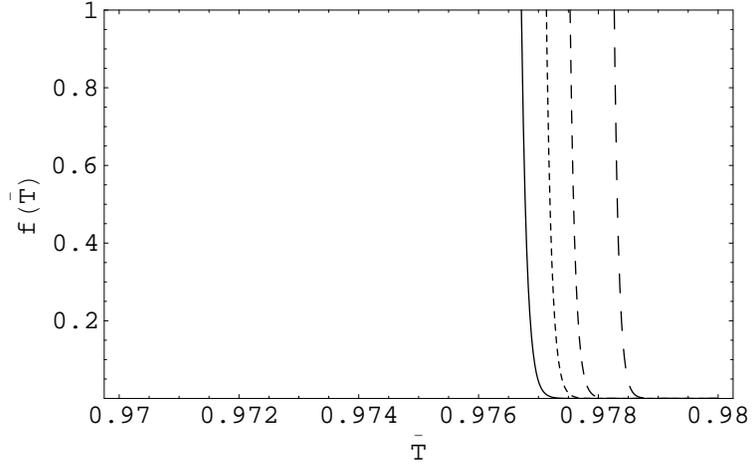}
\caption{Variation of the hadronic fraction $f$ as a function of
the normalized temperature in the small supercooling scenario, for
different values of the brane tension $\lambda $: $\lambda
=5\times 10^8$ MeV$^4$ (solid curve), $\lambda =5\times 10^9$
MeV$^4$ (dotted curve), $\lambda =5\times 10^{10}$ MeV$^4$ (short
dashed curve) and $\lambda =5\times 10^{16}$ MeV$^4$ (long dashed
curve). The bag constant and the critical temperature have the
values: $B^{1/4} = 200$ MeV and $T_c = 125$ MeV, respectively.}
\label{FIG7}
\end{figure}

\begin{figure}[h]
\includegraphics{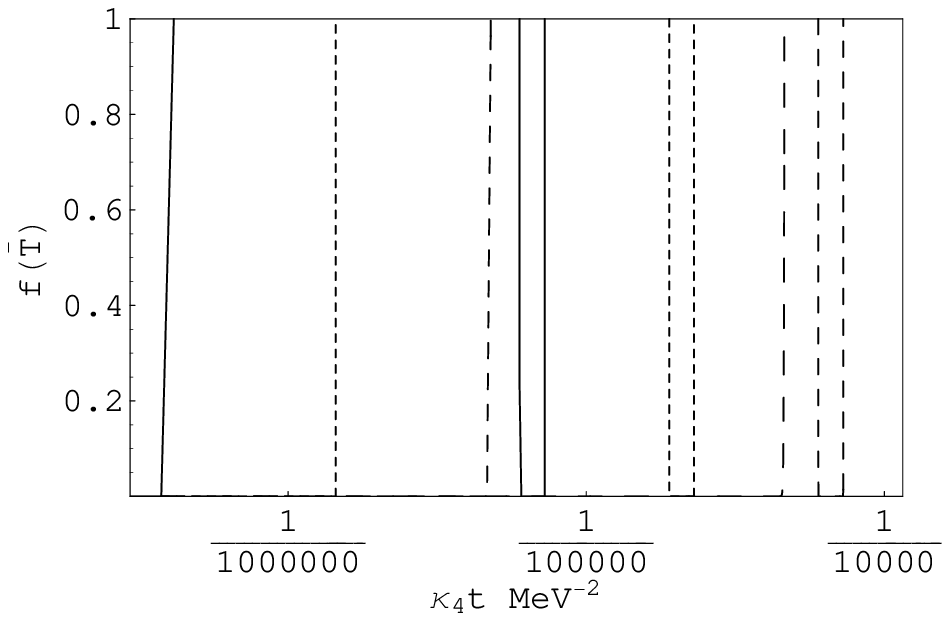}
\caption{ Variation of the hadronic fraction $f$ as a function of
the parameter $\tau =k_4t$ in the small supercooling scenario, for
different values of the brane tension: $\lambda =5\times 10^8$
MeV$^4$ (solid curve), $\lambda =5\times 10^9$ MeV$^4$ (dotted
curve), $\lambda =5\times 10^{10}$ MeV$^4$ (short dashed curve)
and $\lambda =5\times 10^{12}$ MeV$^4$ (long dashed curve). The
bag constant and the critical temperature have the values:
$B^{1/4} = 200$ MeV and $T_c = 125$ MeV, respectively.}
\label{FIG8}
\end{figure}

The plot of $f(\T)$ is presented in Fig. \ref{FIG7} for values of
the parameters used previously, and setting the front velocity
$v_{fr} = 10^{-3}$ and the other constants $w_0,p_0 = 1$. The
fraction of hadronic matter stays very close to zero until the
supercooling temperature is between $\T = 0.98$ and $\T = 0.97$,
then it jumps to $1$ very rapidly. The same behavior can be
verified in the time dependent plot of Fig. \ref{FIG8}. The
comparison between Fig. \ref{FIG8} and Fig. \ref{FIG3} shows
clearly the difference between the first order phase transition
and the supercooling. In the former, the hadronic fraction grows
smoothly over a decade from zero to one, while in the latter it
changes dramatically and almost instantaneously. Physically, what
happens is that, at a certain temperature below the critical
value, an enormous amount of hadronic bubbles are nucleated almost
everywhere in the plasma, which grow explosively to transform all
the plasma into hadrons. The picture is similar to what happen in
standard cosmology \cite{Kajantie:1986hq}, in which the small
supercooling phase transition is also very rapid with respect to
the simple first order phase transition, at a temperature ($T
\simeq 0.98 T_c$), which is very similar to the one we have
obtained. Another remarkable feature that is present in the
supercooling scenario is the important dependence of the
transition temperature on the brane tension, about $10\%$
difference between the lowest and the highest, which eventually
leads to a different time at which the transition occurs.


\section{Discussions and final remarks}\label{SecV}

In the context of brane-world scenarios, in the high density
cosmological phase the Friedmann equation contains deviations to
the $4D$ case, which imposes fundamental phenomenological
consequences on the cosmological evolution, and in particular on
the cosmological phase transitions. In this work, we have analyze
the evolution of the relevant physical quantities, namely, the
energy density, temperature and scale factor before, during and
after the phase transition. It is important to emphasize that in
the early universe the energy density is extremely high, so that
one can neglect the terms linearly proportional to the energy
density with respect to the quadratic terms. This approximation
has been considered in detail throughout this work. In the high
density regime the Hubble function is proportional to the energy
density of the cosmological matter, which drastically affects the
cosmological dynamics of the universe. Moreover, in the early
universe the dark radiation term, the projection of the Weyl
tensor from the bulk, which appears in form of a radiation-like
term in the field equations, may also play an important role. The
magnitude of the brane world effects can be characterized by the
numerical value of the brane tension  $\lambda $. In the limit
$\lambda \to \infty$ we recover standard general relativity
\cite{Ma01}. On the other hand, the study of the quark-hadron
phase transition is also very important from an observational
point of view, since the inhomogeneities generated at the QCD
phase transition might have a noticeable effect on nucleosynthesis
\cite {IgKaKuLa94a}.

By fully taking into account the brane world effects we have found
that the temperature evolution of the universe in the brane world
scenario is different from the idealized standard FRW model. The
temperature of the early universe in the quark phase is smaller in
the brane world scenario, as one can see from Fig.~\ref{FIG1},
where the long dashed curve corresponds (approximately) to the
general relativistic limit. Hence a small value of the brane
tension would significantly reduce the temperature of the
quark-gluon plasma, and accelerate the phase transition to the
hadronic era. An increase of the dark radiation term for fixed bag
constant $B$ and brane tension $\lambda $ gives the same effect,
as one can see from Fig.~\ref{FIG2}. Once the quark-hadron phase
transition starts, the hadron fraction $h$ is again strongly
dependent on the brane tension. For small values of $\lambda $,
$h(t)$ is much higher than in the standard general relativity, as
one can see from Fig.~\ref{FIG3}. The increase of the dark
radiation on the brane also strongly accelerates the formation of
the hadronic phase (see Fig.~\ref{FIG4}), and decreases the time
interval necessary for the transition. A small brane tension and a
high energy density of the dark radiation also tend to reduce the
temperature of the hadronic fluid. Of course, the temperature
evolution also depends upon the relativistic degrees of freedom in
the equation of state and upon the equation of state.
In addition to this, by assuming that the phase transition may be
described by an effective nucleation theory, we have also
considered the case where the Universe evolved through a mixed
phase with a small initial supercooling, and monotonically growing
hadronic bubbles. It was shown that at a certain temperature below
the critical value, an enormous amount of hadronic bubbles are
nucleated, which grow explosively to transform all the plasma into
hadrons, indicating that the small supercooling phase transition
is very rapid with respect to the simple first order phase
transition.

Many details of the QCD phase transition are not yet conclusively
understood. Even the order of transition is still a matter of
debate. An advance in the understanding of the numerical values of
the QCD coupling constants would be very helpful in obtaining
accurate cosmological conclusions. Such an advance may also
provide a powerful method for testing on a cosmological scale the
theoretical predictions of the brane world models and the possible
existence of the extra-dimensions.

\section*{Acknowledgments}

We would like to thank the anonymous referee, whose comments and
suggestions helped us to significantly improve the manuscript. The
work of GDR is supported by I.N.F.N. The work of TH is supported
by the RGC grant HKU 702507P of the government of the Hong Kong
SAR. FSNL was funded by Funda\c{c}\~{a}o para a Ci\^{e}ncia e
Tecnologia (FCT)--Portugal through the grant SFRH/BPD/26269/2006.



\end{document}